\documentclass[aps,preprint]{revtex4}
\usepackage{latexsym}
\begin{document}
\title{\Large \bf Gravitational Field Equations on and off a 3-Brane
World}
\author{\large A. N. Aliev}
\affiliation{ Feza G\"ursey Institute, P.K. 6  \c Cengelk\" oy,
81220 Istanbul, Turkey}
\author{\large A. E. G\"umr\"uk\c{c}\"uo\~glu}
\affiliation{ITU, Faculty of Sciences and Letters, Department of
Physics, 34469 Maslak, Istanbul, Turkey}
\date{\today}

\begin{abstract}
The effective gravitational field equations on and off a $3$-brane world
possessing a $Z_{2}$ mirror symmetry and embedded in a five-dimensional
bulk spacetime with cosmological constant were derived by
Shiromizu, Maeda and Sasaki (SMS) in the framework  of the Gauss-Codazzi
projective approach with the subsequent specialization to the Gaussian
normal coordinates in the neighborhood of the brane. However,
the Gaussian normal coordinates imply a very special slicing of
spacetime and clearly, the consistent analysis of the brane dynamics
would benefit from complete freedom in the slicing of spacetime,
pushing the layer surfaces in the fifth dimension at any rates of
evolution and in arbitrary positions. We generalize the SMS effective
field equations on and off a $3$-brane to the case where there is
an arbitrary energy-momentum tensor in the bulk.
We use a more general setting to allow for acceleration
of the normals to the brane surface through the lapse function
and the shift vector in the spirit of Arnowitt, Deser and Misner.
We show that the gravitational influence of the bulk spacetime on the brane
may be described by a traceless second-rank tensor  $\,W_{ij}\,$,
constructed from the "electric" part of the bulk Riemann tensor.
We also present the evolution equations for the tensor $\,W_{ij}\,$, as well
as for the corresponding "magnetic" part of the bulk curvature.
These equations involve the terms determined by both the nonvanishing
acceleration of normals in the nongeodesic slicing
of spacetime and the presence of other fields in the bulk.
\end{abstract}
\maketitle

\section{Introduction}
The idea of a $3$-brane world is strongly motivated by
developments in string theory. String theory predicts the
existence of higher-dimensional topological defects, D-branes,
which are inherent in the theory and defined by the property that
open string endpoints can live on them \cite{pol}. The attractive
feature of the branes is that they have open string modes confined
to them. In other words, open string modes which involve gauge
fields, fermions and scalars of the Standard Model sector are
trapped on the brane, while closed string modes of the
gravity sector can propogate in higher dimensions. Another
scenario of the localization of gauge fields arises in the
framework of the Horava-Witten model \cite{hw}. In this model the
strongly coupled ten-dimensional $ E_{8}\times E_{8} $ heterotic
string has been related to an eleven-dimensional theory
(M-theory) compactified on an $S^{1}/Z_{2}$ orbifold with gauge
fields propagating on two ten-dimensional branes located on the
boundary hyperplanes with a  $Z_{2}$ mirror symmetry. It has also
been shown that there exists a subsequent
compactification of this model on a deformed Calabi-Yau space,
which leads to a five-dimensional spacetime with boundary
hyperplanes becoming two 3-branes \cite{w,lukas}. The two
3-branes carry the matter and gauge fields on them and are
identified with four-dimensional worlds. In this sense, one may
phenomenologically think of a regime of five-dimensional bulk
Universe which envelopes our physical four-dimensional Universe
residing on a $3$-brane.

Earlier, the idea that our physical Universe might reside
on a $(3+1)$-dimensional hypersurface embedded in a higher
dimensional space was suggested and studied in the framework
of simple field-theoretical models not necessarily connected
with string theory (see Refs.(\cite{rsh}-\cite{gw}). However, a
new striking feature of this idea has emerged in
Arkani-Hamed-Dimopoulos-Dvali (ADD) model \cite{ADD}, where
by combining the $3$-brane universe idea with
the original idea of the Kaluza-Klein compactification, the authors
suggested to solve the long-standing hierarchy problem of high-energy
physics. The ADD model is built on the basic ingredients that our
observable four-dimensional spacetime is a $3$-brane with all Standard
Model fields localized on it, while gravity is free to propagate in all
extra spatial dimensions. The size of the extra dimensions may be much
larger than the conventional Planckian length ($\sim 10^{-33}
cm $). Thus, in contrast to the original Kaluza-Klein scenario,
in the ADD models the extra dimensions are supposed to
manifest themselves as physical ones. One of the dramatic
consequences of this model is that the fundamental scale of
gravity might become as low as the ultraviolet scale of the Standard Model
(of the order of a few TeV). This also raises the idea of
TeV-size mini black holes and their detectability in cosmic ray
experiments or at future high energy colliders \cite{emp1}-\cite{dl}.

An alternative approach to solution of the hierarchy problem was
put forward by Randall and Sundrum, who proposed a new
higher-dimensional model consisting of a single extra spatial
dimension \cite{rs1}. This model (RS1) is based on two 3-branes that are
located at the boundaries of an orbifold $S^{1}/Z_{2}$ in a five
dimensional anti-de Sitter $(AdS_5)$ spacetime. Under a certain
fine balance between the branes' self-gravity and bulk cosmological
constant, the ultraviolet scale is generated from the large
Planckian scale through an exponential warp function of a small
compactification radius. Randall and Sundrum also proposed
the second model (RS2) suggesting that our observable Universe
resides on a single self-gravitating 3-brane with positive
tension, which is embedded in $ AdS_5$ bulk spacetime
with infinite extra dimension and negative cosmological
constant \cite{rs2}. In this model the fine
tuning between the brane and the bulk leads to the phenomenon of
localization on the $3$-brane of a $5$-graviton zero mode along with
rapid damping at large distances of massive KK modes. In other words,
the RS2 model supports the conventional potential of Newtonian
gravity on the $3$-brane with high enough accuracy.

Due to their striking phenomenological consequences the ADD and
RS braneworld models have caused enormous activity
in particle physics, gravity and cosmology, paving the way to
braneworld "scenarios" in these fields. A comprehensive description of
these scenarios can be found in recent reviews \cite{csaki}-\cite{rubakov}.
Among them, the study of the properties of gravity localized
around the branes is of particular interest. In \cite{gt}-\cite{gkr}
it has been shown that not only the ordinary Newtonian potential, but
also four-dimensional Einstein gravity is localized around a $3$-brane
in RS2 braneworld, if in the linearized approach one properly takes
into account the so-called brane-bending effect. A nonlinear model
of plane wave propagating along the brane was studied in \cite{chgw}.
The subsequent analysis of this issue has been performed
in \cite{sms1}-\cite{sms2}, where the authors handled the problem
in a nice covariant way, deriving the effective gravitational field
equations on a $3$-brane embedded in a five-dimensional bulk spacetime
with $Z_{2}$-symmetry. These equations involve the higher-dimensional
corrections to the ordinary Einstein equations, which are
determined by a term quadratic in the energy-momentum tensor of the brane,
as well as a curvature term from the bulk space. These correction terms
become ignorable in the low energy limit.
Further generalizations of this approach to include the dilaton fields
in the bulk, as well as high-order curvature terms of Gauss-Bonnet
combinations have been given in \cite{mw}-\cite{mt}. The case of null
radiation in the bulk was considered in \cite{roy}.

The basic idea of the paper \cite{sms1} and the other related
works was to use the Gauss-Codazzi projective approach and
project the field equations in the bulk on a $3$-brane
with the subsequent specialization to the Gaussian normal
coordinates in the neighborhood of the $3$-brane. However,
as is well-known the Gaussian normal coordinates imply a
very special slicing of  spacetime in the sense that the geodesics
orthogonal to a given hypersurface remain orthogonal to all
successive hypersurfaces in the slicing \cite{mtw}. This slicing may
not affect the field equations on the brane, but it will change
the off-brane evolution equations. Clearly, the consistent analysis
of the brane dynamics would benefits from complete freedom in the
slicing of spacetime by timelike hypersurfaces, pushing them in the fifth
dimension at arbitrary rates and in arbitrary positions.

It is the purpose of this paper to present the effective
gravitational field equations on a $3$-brane
in the five-dimensional bulk, as well as the evolution equations
off the brane in the case of arbitrary energy-momentum tensor in the bulk.
We use, instead of the Gaussian normal coordinates, more general
coordinate setting in the spirit of Arnowitt, Deser and
Misner (ADM) \cite{adm}. The use of the ADM type coordinates implies
the slicing of the spacetime by timelike hypersurfaces pierced
by a congruence of spacelike curves that are not geodesics and do
not intersect the hypersurfaces orthogonally. In this approach one
has freedom in analyzing the bulk dynamics in the most general case.
We show that in the framework of this more general geometric setting
the effective gravitational field equations
on the $3$-brane have the same form as obtained earlier
\cite{sms1} in the absence of matter fields in the bulk. We show
that the gravitational influence of the bulk spacetime on the brane
may be described through a traceless second-rank tensor  $\,W_{ij}\,$,
properly constructed from the "electric" part of the bulk Riemann tensor.
We also show the evolution equations in the bulk are
significantly modified due to the acceleration of normals in the
nongeodesic slicing of spacetime and the presence of other fields
in the bulk as well.

The paper is organized as follows. In Sec.II we describe the general
formalism of a $(4+1)$ decomposition of a five-dimensional bulk spacetime
using an approach similar to the ADM approach in the
Hamiltonian formulation of four-dimensional general relativity.
In Sec.III we begin with the decomposition of the connection
coefficients using the usual definition of the Christoffel symbols.
Then we derive the decomposition formulas for the Ricci, Einstein
and Riemann tensors, in accordance with the $(4+1)$ splitting of the
spacetime metric, without appealing to the Gauss-Codazzi projective
approach. The effective field equations on the brane, as well as
the conservation equations on it are given in Sec.IV. Finally, in Sec.V
we derive the evolution equations off the brane for the corresponding
projected parts of the curvature and energy-momentum tensors in the bulk
space.

\section{Geometric Setting}

We consider a five-dimensional bulk spacetime with manifold $M$ and
metric $ g_{AB} $ and suppose that the spacetime includes a $3$-brane,
a $(3+1)$-dimensional hypersurface, which, in turn, is endowed with the
metric $h_{AB}$. Assuming that the bulk spacetime is covered by
coordinates $ x^{A}$ with $A=0,1,2,3,5$ we introduce an arbitrary
scalar function
\begin{equation}
Z= Z(x^A) \,,
\label{scalarf}
\end{equation}
such that $Z$= const describes a layer surface in the family of
non-intersecting timelike hypersurfaces $\Sigma_Z$, filling $M$, with
the $3$-brane located at the hypersurface $Z=0$\,.

The total action corresponding to the above setting consists of the
Einstein-Hilbert action  with bulk matter fields and the brane action
\begin{equation}
S=\frac{1}{2 \kappa_{5}^2}\,\int_{M} \,d^5 x\sqrt{-g}\,
(\,^{(5)}R -2\,\Lambda_5)
+\int_{M} \,d^5 x\sqrt{-g}\, ^{(5)}L_{bulk} + \int_{brane} \,d^4 y \sqrt{-h}
\,L_{brane}\,,
\label{action}
\end{equation}
where $\,\kappa_{5}^2=8\pi G_5\,$, with $\,G_5\,$ being the gravitational
coupling constant, $y^{i}$ with $i=0,1,2,3$ are
coordinates intrinsic to the brane and $\Lambda_{5}$ is the
bulk cosmological constant. To avoid an undesired term in
extremization of this action with respect
to variations of $g_{AB}$ and  $h_{AB}$, the Gibbons-Hawking boundary term
\cite{gh} is also implied in it. The field equations
derived from the action (\ref{action}) have the form
\begin{eqnarray}
^{(5)}G_{AB}&=& \, ^{(5)}R_{AB}- \frac{1}{2}\,g_{AB}\,^{(5)}R
\,=\, - \Lambda_{5} \, g_{AB}+
\kappa_5^2\left(^{(5)}T_{AB}+\sqrt{\frac{h}{g}}\,\,\tau_{AB}\,
\delta(Z)\right)\,,
\label{5dfeq}
\end{eqnarray}
where $^{(5)}T_{AB}$ is the energy-momentum tensor
for the generic fields in the bulk, $\,\tau_{AB}\,$
is the energy-momentum tensor in the braneworld, the $\delta$-function
singularity means that the brane energy-momentum has the form of a thin
sheet distribution and the quantities $\,g\,$ and $\,h\,$ are the metric
determinants of $g_{AB}$ and $h_{AB}$, respectively.

The effective gravitational field equations in the braneworld are obtained
by implementing a $(4+1)$ decomposition of the five-dimensional bulk
spacetime, viewing it in terms of the four-dimensional hypersurface
of a $3$-brane and a spatial fifth dimension.
(In the following we shall basically adopt the notations
of Refs.\cite{Israel}-\cite{eric} and use the reduced on-brane coordinate,
instead of general five-dimensional ones). Clearly, one can
introduce the unit spacelike normal to the brane surface as
\begin{equation}
n_A = N\,\partial_A Z\,,
\label{vnormal}
\end{equation}
satisfying the normalization condition
\begin{equation}
g_{AB}\,n^A\,n^B = 1\,,
\label{normalization}
\end{equation}
where  the scalar function
$$N=\left| g^{AB}\,\partial_A Z\,\partial_B Z \right|^{-1/2}$$
is  called the lapse function. The metric intrinsic to the  brane
can be obtained through defining the infinitesimal displacements
within the brane. We recall the parametric equation of the brane
$ x^A=x^A (y^i) $, that implies the existence of a local frame given by
the set of four vectors
\begin{equation}
e^A_i = \frac{\partial x^A}{\partial y^i}~~~,
\label{tvector}
\end{equation}
which are tangent to the brane. It is clear that they satisfy the
orthogonality condition
\begin{equation}
n_A\,e^A_i = 0 \,.
\label{orthogonal}
\end{equation}
With this local frame the infinitesimal displacements on the brane are
determined by the induced metric
\begin{equation}
h_{ij} = g_{AB} \,e^A_i \,e^B_j\,\,,
\label{indmetric}
\end{equation}
and its inverse, satisfying the completeness relation on the brane
\begin{equation}
h^{ik}\,h_{kj} = \delta_{j}^i\,\,.
\label{completeness}
\end{equation}
From equations (\ref{orthogonal}) and (\ref{indmetric}) it follows that
the bulk spacetime metric can be given as
\begin{equation}
g_{AB} = n_A n_B +h_{ij} e^i_A e^j_B \,\,,
\label{5metric1}
\end{equation}
while the inverse metric has the form
\begin{equation}
g^{AB} = n^A n^B +h^{ij} e_i^A e_j^B\,\,.
\label{5metric}
\end{equation}
In obtaining the last expression we have also used the completeness
relation
\begin{equation}
e_A^i\,e^A_j = \delta_{j}^i \,\,,
\label{orthogonal1}
\end{equation}
for the basis vectors on the brane.

Next, we shall introduce a spacelike vector $Z^A$, which formally can be
thought of as an "evolution vector" into the fifth dimension. Having chosen
the parameter along the orbits of this vector as Z, we can write the
relation
\begin{equation}
Z^A\,\partial_A Z =1\,,
\label{evector1}
\end{equation}
which means that the vector $Z^A$ is tangent to a congruence of curves
intersecting the succesive hypersurfaces in the slicing of spacetime.
We are interested in the general case, when
the curves are not geodesics and they do not intersect the hypersurfaces
orthogonally. The evolution vector $Z^A$ is not necessarily parallel to the
normal vector $n^A$ and it can be decomposed
into its normal and tangential parts
\begin{equation}
Z^A = N\,n^A +N^i\,e^A_i \,\,,
\label{evector2}
\end{equation}
where the four vector $N^i$ is known as the shift vector
(see Ref.\cite{eric}).
With this construction one can always define an alternative coordinate
system $\,(y^i, y^5)\equiv (y^i, Z)\,$ on $M$, and express the
spacetime metric in these new coordinates.
Then the relations (\ref{tvector}) and (\ref{evector1})
imply that
\begin{eqnarray}
e^A_i & = &\left(\frac{\partial x^A}{\partial y^i}\right)_Z\,
= \delta^A_i\,\,, ~~~~~~~~~~
Z^A= \left(\frac{\partial x^A}{\partial Z}\right)_{y^i} = \delta^A_Z \,\,.
\label{ntevectors}
\end{eqnarray}
Furthermore, we note that
\begin{eqnarray}
dx^A &=& Z^A dZ +e^A_i dy^i
= \left(N dZ \right) n^A + e^A_i \left(dy^i+N^i dZ \right)\,,
\label{diff}
\end{eqnarray}
and the spacetime interval is decomposed into the form
\begin{eqnarray}
ds^2&=&g_{AB} dx^{A}dx^{B} \nonumber \\
&=& h_{ij}\, dy^i\,dy^j + 2N_i \,dy^i \,dZ +
\left(N^2 +N_i N^i \right)\,dZ^2 \,.
\label{demetric}
\end{eqnarray}
It follows that the components of the five-dimensional metric
can be written in terms of the induced 4-metric, the lapse and
the shift functions as
\begin{equation}
g_{AB} =
\begin{pmatrix}
{h_{ij}  &&  N_i \cr\cr  N_j && N^2 +N_i N^i}
\end{pmatrix} \,\,.
\label{redmetric}
\end{equation}
For the metric determinant we have
\begin{equation}
\sqrt{-g} = N \sqrt{-h}\,,
\label{determinant}
\end{equation}
while the inverse metric takes the form
\begin{equation}
g^{AB} =
\begin{pmatrix}
{h^{ij} + N^i N^j/N^2 && - N^i/N^2 \cr\cr - N^j/ N^2
 && 1/N^2}
\end{pmatrix} \,\,.
\label{invmetric}
\end{equation}
We also note that the covariant components of the spacelike unit
normal vector in equation (\ref{vnormal}) are given by
\begin{equation}
n_A =\left(0,0,0,0,N \right)\,,
\label{normaldown}
\end{equation}
and its contravariant components are obtained through the use of equations
(\ref{normalization}) and (\ref{redmetric}). We have
\begin{equation}
n^A = \left( -\frac{N^i}{N}\,\,,\,\,\, \frac{1}{N} \right)\,.
\label{normalup}
\end{equation}

Finally, to describe the bending of the brane surface in the bulk
we shall introduce the extrinsic curvature of the brane \cite{eric},
calculating the derivative of the normal vector as
\begin{equation}
\nabla_A\,n_B= K_{AB}+n_A a_B\,,
\label{acceleration1}
\end{equation}
where $\,\nabla\,$ is the covariant derivative operator associated
with the bulk metric $\,g_{AB}\,$,  the extrinsic curvature
tensor $\,K_{AB}\,$ is a symmetric tensor tangential to the brane
$(\,K_{AB} n^A=0 \,)$ and the 5-acceleration of the normal vector
is given by
\begin{equation}
a_A = n^B \,\nabla_B\,n_A\,\,,
\label{acceleration}
\end{equation}
which with equations (\ref{ntevectors}) and (\ref{normalup})
can be written in the $\,(4+1)\,$ component form
\begin{eqnarray}
a_A = (a_i, a_5) &=&
-\frac{1}{N}\,\left(D_i N\,\,,\,\,\,N^i\,D_i N \right)\,\,.
\label{acccomp}
\end{eqnarray}
Here $D$ is the covariant derivative operator defined with respect
to the brane metric $\,h_{ij}\,$ and it is clear that $\,a_A n^A=0\,$.

The projected extrinsic curvature tensor is the symmetric 4-tensor
given by
\begin{eqnarray}
K_{ij}&=& \nabla_{(B}n_{A)}\, e^A_i e^B_j\,\
=\frac{1}{2}\left({\pounds_n} g_{AB}\right) e^A_ie^B_j\,\,,
\label{excurv1}
\end{eqnarray}
where the symbol $ \pounds$ stands for the Lie derivative.
The extrinsic curvature can be related to the evolution of the brane metric
into the fifth spatial dimension through its Lie derivative
along the evolution vector $Z^A$. We obtain that
\begin{equation}
K_{ij}=\frac{1}{2N}\left[\left({\pounds_Z} g_{AB}\right) e^A_i
e^B_j-D_i N_j-D_j N_i\right] \,\,,
\label{excurv2}
\end{equation}
which, with equations in (\ref{ntevectors}), reduces to the form
\begin{equation}
K_{ij}=\frac{1}{2N} \left( \partial_{5} h_{ij}- D_i N_j-D_j N_i\right)\,.
\label{excurv3}
\end{equation}
Here and in what follows $\,\partial_5= \partial/\partial{Z}\,\,$ and
the indices of all $4$-tensors intrinsic to the brane are raised
and lowered with respect to the metric $h_{ij}$.

\begin{center}
\section{Decomposition of Connection Coefficients and Curvature Tensors}
\end{center}

In accordance with a $\,(4+1)\,$ decomposition of the five
dimensional spacetime metric given in (\ref{demetric}), we now
need to establish the reduction formulas between the five
dimensional connection coefficients and curvature tensors and the
quantities characterizing the intrinsic and extrinsic properties
of the brane surface. We start with the usual definition of the
Christoffel symbols in five dimensions
\begin{equation}
\Gamma^A_{BC}=\frac{1}{2}\, g^{AD}\left(\partial _B
g_{CD}+\partial _C g_{BD}- \partial _D g_{BC}\right)\,,
\label{5chr}
\end{equation}
and substituting into it the metric decompositions given in
(\ref{redmetric}) and (\ref{invmetric}), we obtain that the
nonvanishing components of the Christoffel symbols are split into
the form
\begin{eqnarray}
\Gamma^5_{ij} & = & - \frac{1}{N}K_{ij}\,\,, \nonumber\\
\Gamma^5_{i5}& = & N^j\,\Gamma^5_{ij} +\frac{1}{N}\,D_{i} N\,\,,\nonumber\\
\Gamma^i_{jl}& =&\lambda^i_{jl}-N^i\,\Gamma^5_{jl}\,\,,\nonumber\\
\Gamma^i_{j5}& = & -N^i\,\Gamma^5_{5 j}+ N \,K^i_j +D_j \,N^i
\,\,,\nonumber\\
\Gamma^5_{55} & = & N^i\, \Gamma^5_{i5} +
\frac{1}{N}\,\partial_{5} N\,\,, \label{redchr}
\end{eqnarray}
$$\Gamma^i_{55} = - N^i\, \Gamma^5_{55} +\partial_5 \,N^i + N^j \,D_j \,N^i
+ N\,(2\, K^i_j \,N^j - D^i N) \,\,,$$
where the quantities
\begin{equation}
\lambda^i_{jl}=\frac{1}{2}\,h^{im}\left(\partial _j
h_{lm}+\partial _l h_{jm}- \partial _m h_{jl}\right)\,,
\label{4chr}
\end{equation}
are the Christoffel symbols on the brane formed from the metric
$\,h_{ij}\,$.

The decomposition of the five dimensional Ricci tensor can be
obtained through the use of the above reduction formulas
(\ref{redchr}) and (\ref{determinant}) in the general expression
\begin{equation}
^{(5)}R_{AB} = \partial_C \Gamma^C_{AB} -
\partial_A \partial_B \left(\ln{\sqrt{-g}} \right) + \Gamma^C_{AB}
\partial_C \left(\ln{\sqrt{-g}} \right) - \Gamma^C_{AD}
\Gamma^D_{BC}\,. \label{ricci}
\end{equation}
After performing straightforward calculations we find that
\begin{eqnarray}
^{(5)}R_{i j} & = & R_{ij} - \frac{1}{N}\,\left[\left(\partial_5
- \pounds_{\vec{N}}\right) K_{ij} + D_i  D_j N\right] +2\,K^m_i
K_{mj} - K K_{ij} \,,
\nonumber \\[3mm]
^{(5)}R_{5 i} & = & ^{(5)}R_{ij} \,N^j - N \left(D_i \,K-D_j
\,K^j_i \right)\,,
\nonumber \\[3mm]
^{(5)}R_{55} & = & ^{(5)}R_{ij} \,N^i N^j - N \left(\partial_5 \,K
+ \Box N \right) - N^2 \,K_{lm} \,K^{lm}
\nonumber\\
&& + \,2 N \,N^j\,  D_i \left(K^i_j - \frac{1}{2 }\,\delta^i_j \,K
\right)\,,
\label{redricci}
\end{eqnarray}
where $\,R_{ij}\,$ is the Ricci curvature of the braneworld,
$\,K=\,h^{ij}\,K_{ij}\,$ is the trace of the extrinsic curvature,
the operator $\,\Box = D_m\,D^m\, $ is the D'Alembertian
acting on the brane and the Lie derivative  $\pounds_{\vec{N}}$
is taken along the shift vector $\,N^i\,$.

Similarily, for the five dimensional Ricci scalar we obtain the
following reduction formula
\begin{equation}
^{(5)}R = R - \frac{2}{N}\left(\partial_5\, K -
\pounds_{\vec{N}}\,K + \Box N \right) - K_{lm}\,K^{lm} - K^2\,\,,
\label{redscalar}
\end{equation}
with the four-dimensional  Ricci scalar $\,R\,$ defined on the
brane. We note that in obtaining the above expressions we have
repeatedly used the following useful relations
\begin{eqnarray}
\partial_5 \,h^{ij}  & = & - h^{im} \,h^{jl}\,\partial_5 h_{lm}\,\,,
\nonumber \\[2mm]
\partial_5 \ln \sqrt{-h} & = &  N\,K + D_i\,N^i \,\,,
\nonumber\\[2mm]
h^{ij}\,\partial_5 \,K_{ij} & = & \partial_5 \,K +
2\,\left(K^i_j\,D_i\,N^j + N\, K_{ij}\,K^{ij} \right) \,\,.
\label{useful}
\end{eqnarray}

We can now  establish the reduction formulas for the components of
the Einstein tensor in equation (\ref{5dfeq}). Using equations
given in (\ref{redricci}), (\ref{redscalar}) and taking into
account relations (\ref{useful}) we obtain
\begin{eqnarray}
^{(5)}G_{ij} & = & G_{ij} - \frac{1}{N} \left[ \left(\partial_5 -
\pounds_{\vec{N}} \right) \left(K_{ij} - h_{ij}\,K\right) +D_i D_j
N \right] - 3\,K K_{ij} + 2\,K^m_i K_{mj}
\,\nonumber\\
&&  + \frac{1}{2}\,h_{ij} \left(K^2 + K_{lm} K^{lm} +
\frac{2}{N}\,\Box N \right)\,,
\nonumber \\[3mm]
^{(5)}G_{i5} & = & ^{(5)}G_{ij}\,N^j - N\left(D_i K-D_j
K^j_i\right)\,,
\nonumber \\[3mm]
^{(5)}G_{55} & = & ^{(5)}G_{ij}\, N^i N^j -
\frac{1}{2}\,N^2\left(R-K^2+K_{lm} K^{lm}\right) +2\,N N^j
D_i\left(K^i_j-\delta^i_j \,K\right) \,\,, \label{redeinstein}
\end{eqnarray}
where $\,G_{ij}\,$ is the Einstein $4$-tensor in the braneworld.

We shall also need the decomposition formulas for the Riemann
curvature tensor of the metric (\ref{demetric}). They can be
obtained from the general expression
\begin{equation}
^{(5)}R^A_{\,\,\,BCD}=\partial_C\Gamma^A_{BD} - \partial_D
\Gamma^A_{BC}+ \Gamma^M_{BD}\Gamma^A_{CM}
-\Gamma^M_{BC}\Gamma^A_{DM} \,\, \label{griemann}
\end{equation}
by substituting into it the decomposition of the Christoffel
symbols given in equation (\ref{redchr}). Having done this, for
the covariant components of the five dimensional Riemann curvature
we  find
\begin{eqnarray}
^{(5)}R_{ijlm} & = & R_{ijlm} + K_{im} K_{jl}-K_{il} K_{jm}\,\,,
\nonumber \\ [3mm] ^{(5)}R_{5ijl} & = & ^{(5)}R_{mijl}\,N^m -
N\left(D_j K_{il}-D_l K_{ij}\right)\,\,, \nonumber \\ [3mm]
^{(5)}R_{5i5j} & = & ^{(5)}R_{5imj}\,N^m - N \left[\,
\left(\partial_5 - \pounds_{\vec{N}} \right) K_{ij}
+D_i D_j N \,\right] \nonumber \\
&& +\,N \left[\,N^m\left(D_i K_{mj}-D_m K_{ij} \right) + N K^m_i
K_{mj}\,\right]\,\,, \label{redriemann}
\end{eqnarray}
where the unmarked Riemann tensor is intrinsic to the brane
surface. In the derivation of these expressions, in addition to
equations in (\ref{useful}), we have also used the relation
\begin{equation}
\partial_5\,\lambda^i_{jl}= R^i_{\,\,\,lmj} \,N^m +D_j D_l N^i +
D_j (N\,K^i_l) + D_l (N\,K^i_j) - D^i (N\,K_{jl})\,\,.
\label{5lam}
\end{equation}
It is important to note that among equations in (\ref{redriemann})
only the last equation involves the term describing the evolution
into the fifth dimension. We recall that the first equation is
known as the Gauss equation. In the case of vanishing acceleration
of the normals $\,(N=1,\,\, N^i=0 )\,$, the above formulas are
in agreement with those obtained in \cite{mt} within
the Gauss-Codazzi projective approach.

\vspace{5mm}
\section{Effective Equations on The Brane}

We are now in position to write down the effective gravitational
field equations on a $3$-brane substituting the reduction formulas
(\ref{redeinstein}) into the basic equation (\ref{5dfeq}) together
with a suitable decomposition of its right-hand-side. First, we
note that
\begin{eqnarray}
^{(5)}G_{i5}&=&{}^{(5)}G_{iA}\,Z^A
= {}^{(5)}G_{ij}N^j+N~{}^{(5)}G_{iA}n^A\,\,,\\[3mm]
\label{einstein5i}
^{(5)}G_{55}&=&{}^{(5)}G_{AB}Z^A Z^B\,\nonumber\\
&=&{}^{(5)}G_{ij}\,N^i\,N^j+2 N ~^{(5)}G_{iA}\,N^i\,n^A  +
N^2~^{(5)}G_{AB}\,n^A \,n^B \,\,, \label{einstein55}
\end{eqnarray}
and comparing these expressions with the second and third
equations in (\ref{redeinstein}) we arrive at equations
\begin{equation}
\frac{1}{2}\,\left(R-K^2 + K_{lm}K^{lm}\right) =\,\Lambda_{5}
-\kappa_5^2~P \,\,, \label{constr1}
\end{equation}
\begin{equation}
D_m K^m_i-D_i K = \kappa_5^2~ J_i\,\,, \label{constr2}
\end{equation}
where  $\,P=\ ^{(5)}T_{AB}\,n^A n^B\,$ is a sort of "pressure"
exerted on the brane, while the $4$-vector
$\,J_i=\ ^{(5)}T_{iB}n^B\,$ describes the energy-momentum flux onto
or from the brane. Using the analogy with the Hamiltonian
formulation of general relativity \cite{adm} we can think of
these equations as the Hamiltonian constraint and the momentum
constraint equations on the brane, respectively.

Next, we insert the remaining first equation in
(\ref{redeinstein}) into equation (\ref{5dfeq}) and rewrite it in
form
\begin{eqnarray}
G_{ij}& - &\frac{1}{N} \left[ \left(\partial_5 - \pounds_{\vec{N}}
\right) \left(K_{ij} - h_{ij}\,K\right) +D_i D_j N \right] - 3\,K
K_{ij} + 2\,K^m_i K_{mj}
 \,\nonumber\\
 & + &\,\frac{1}{2}\,h_{ij} \left(K^2 + K_{lm} K^{lm} +
\frac{2}{N}\,\Box N \right)
= - \Lambda_{5} \, h_{ij}+
\kappa_5^2\left(^{(5)}T_{ij}+\sqrt{\frac{h}{g}}\,\,\tau_{ij}\,
\delta(Z)\right)\,. \label{beq1}
\end{eqnarray}
In order to assign to this equation a physical meaning on the
brane we must do two things. First, we must relate the
$\delta$-function behaviour in the brane energy-momentum tensor on
the right-hand-side of equation (\ref{beq1}) to the jump in the
extrinsic curvature of the brane on its evolution into the fifth
dimension. Further, we must express the term describing the
evolution into the extra dimension in terms of four-dimensional
quantities on the brane, more precisely in terms of their limiting
values on the brane. At this stage, it is useful to impose on the
brane $Z_2$-symmetry and integrate with an appropriate weight the
equation (\ref{beq1}) across the brane surface along the orbits of
the evolution vector $\,Z^A\,$. Then passing to the zero limit
$\,Z\rightarrow \pm \,0\,$ and assuming that the quantities
$\,K^{+}_{ij} \,$ and $\,K^{-}_{ij}\,$ evaluated on both sides of
the brane, respectively, remain bounded, we arrive at Israel's
junction condition \cite{Israel}
\begin{eqnarray}
\left[h_{ij}\right]_{\pm}=h^+_{ij}- h^-_{ij}&=&0\,,\nonumber\\
\left[K_{ij}\right]_{\pm} - h_{ij} \left[K \right]_{\pm}
&=&-\kappa_{5}^2\,\tau_{ij}\,. \label{jcisrael}
\end{eqnarray}
Since we have imposed on the brane $Z_2$-symmetry, that implies
$\,Z^A\rightarrow -Z^A\,\;\;\;\left(n^A\rightarrow -n^A \right)$
and hence,
\begin{equation}
K^+_{ij}= - K^-_{ij}\,. \label{Z2sym}
\end{equation}
Taking this into account in equation (\ref{jcisrael}) and omitting
the indices $\,\pm\,$ we obtain
\begin{equation}
K_{ij}=-\frac{1}{2}\,\kappa_5^2\,\left(\tau_{ij} -
\frac{1}{3}\,h_{ij}\,\tau\right)\,, \label{brextcur}
\end{equation}
where $\,\tau=\tau_{ij}\,h^{ij} \,$ is the trace of the
energy-momentum tensor of the brane.

Next, we note that
\begin{equation}
^{(5)}R_{5i5j}={}^{(5)}R_{ABCD}\,Z^A Z^C\,e^B_i\,e^D_j\,,
\label{riemann5i5j}
\end{equation}
which by means of equation (\ref{evector2}) and the second
equation in (\ref{redriemann}) can be written in the form
\begin{equation}
^{(5)}R_{5i5j}=N^2 A_{ij}+ {}^{(5)}R_{5imj}\,N^m - N
B_{jim}\,N^m\,, \label{empart}
\end{equation}
where we have introduced the quantities
\begin{eqnarray}
A_{ij}&=&{}^{(5)} R_{ABCD}\,n^A \,n^C e^B_i e^D_j \,,\nonumber\\
B_{jim}&=& {}^{(5)} R_{ABCD}\,n^A e^B_j e^C_i e^D_m =
2\,D_{[m}K_{i]j}\,\,, \label{defab}
\end{eqnarray}
which, by analogy with their corresponding counterparts in ordinary
general relativity \cite{landau}, can be thought of as the
"electric"  part and the "magnetic" part of the bulk Riemann tensor,
respectively. Comparing equation (\ref{empart}) with the last equation in
(\ref{redriemann}) we find that
\begin{equation}
-\,\frac{1}{N}\left(\partial_5 K_{ij}-\pounds_{\vec{N}}K_{ij}
\right) = A_{ij} - K^m_i K_{mj} - D_i a_j +a_i \,a_j\,\,,
\label{excev1}
\end{equation}
where we have inserted the four-dimensional acceleration
$\,a_i= a_A e_i^A\,$ given in  (\ref{acccomp}).
Similarily, from the reduction formula (\ref{redscalar}) for the
Ricci scalar, it follows that
\begin{equation}
\frac{1}{N} \left(\partial_5 - \pounds_{\vec{N}}\right)
K=\frac{1}{2} \left(R-{}^{(5)}R-K_{lm} K^{lm}- K^2 \right) + D_m
a^m - a^2 \,\,, \label{excev2}
\end{equation}
where $\,a^2= a_m a^m\,$. Taking into account the obvious relation
\begin{equation}
\frac{1}{N}\left(\partial_5-\pounds_{\vec{N}}\right) K h_{ij} =
\frac{1}{N}h_{ij} \left(\partial_5 K -\pounds_{\vec{N}}
K\right)+2\,K K_{ij}\,, \label{excev3}
\end{equation}
and combining equations (\ref{excev1}) and (\ref{excev2}) we
arrive at desired expression relating the evolution term in
(\ref{beq1}) to the four-dimensional quantities on the brane
\begin{eqnarray}
-\frac{1}{N}\left(\partial_5 - \pounds_{\vec{N}}\right)
\left(K_{ij} - h_{ij} K\right) & = & A_{ij} + \frac{1}{2}\,h_{ij}
\left(R- K_{lm} K^{lm} - K^2 -{}^{(5)}R \right) -K^m_i\,K_{mj}\,\
\nonumber\\
&&+2\,K K_{ij} - D_i\, a_j + a_i \,a_j +h_{ij} (D_m a^m -
a^2)\,\,. \label{exev3}
\end{eqnarray}
Inserting this expression into (\ref{beq1}) and approaching the
brane from either $\,+\,$ or $\,-\,$ side, we obtain the effective
gravitational field equations on the brane in the form
\begin{eqnarray}
G_{ij}&+&\frac{1}{2}\,h_{ij}\left(K^2-K_{lm}K^{lm}\right)+K^m_i
K_{mj} - K K_{ij} + A_{ij}\,\nonumber\\ &=& -\frac{1}{3}\,h_{ij}
\Lambda_5 + \kappa_5^2\left[^{(5)}T_{ij}+
h_{ij}\left(P-\frac{1}{3}\,^{(5)}T\right)\right] \,\,,
\label{breq2}
\end{eqnarray}
where we have used equation (\ref{constr1}), while the extrinsic
curvature terms are determined through equation (\ref{brextcur}).

It may be useful to introduce a traceless tensor $\,W_{ij}\,$,
constructed from the electric part of the Riemann curvature
according to the relation
\begin{equation}
W_{ij}=A_{ij}-\frac{1}{4}\,h_{ij}\, A \label{wtrl}\,\,,
\end{equation}
where $\,A\,$ is the trace of $\,A_{ij}\,$ with respect to the induced
metric  $\,h_{ij}\,$. We have
\begin{equation}
A=\frac{1}{2}\,\left(^{(5)}R-R-K_{lm}K^{lm}+K^2\right)\,.
\label{atrl}
\end{equation}
Substituting this new tensor into equation (\ref{breq2}) and
expressing the extrinsic curvature terms through the
energy-momentum tensor in the braneworld, we obtain the alternative
form of the effective field equations
\begin{equation}
G_{ij}= - \frac{1}{2} h_{ij}\, \left(\Lambda_5 - \kappa_5^2\,P
\right) - W_{ij} -3 \kappa_5^2 \,U_{ij} - \kappa_5^4\,\tilde
T_{ij}\,\,, \label{breq3}
\end{equation}
where the source term $\,\tilde T_{ij}\,$ is quadratic in
the energy-momentum tensor of the brane and given by
\begin{equation}
\tilde T_{ij} =  \frac{1}{4}\, \left[ \tau^m_i \tau_{mj} -
\frac{1}{3}\,\tau \tau_{ij} - \frac{1}{2}\,h_{ij}
\left(\tau_{lm}\tau^{lm} - \frac{1}{3}\,\tau^2\right)\right]\,\,,
\label{quadratic}
\end{equation}
while
\begin{equation}
U_{ij}= - \frac{1}{3}\,\left(^{(5)}T_{ij} - \frac{1}{4}\,h_{ij}
h^{lm}\,^{(5)}T_{lm}\right)\,\,, \label{utrl}
\end{equation}
is the traceless, with respect to the metric $\,h_{ij}\,$,
projection of the bulk energy-momentum tensor onto the brane. On
the other hand, taking into account the decomposition of the
Riemann tensor in five dimensions in terms of the traceless Weyl
tensor $\, C_{ABCD}\,$ and the Ricci tensor
\begin{equation}
R_{ABCD}=C_{ABCD}+\frac{2}{3}\,\left(g_{A[C}R_{D]B}-g_{B[C}R_{D]A}\right)
-\frac{1}{6}\,R~g_{A[C}g_{D]B}\,\,,
\label{weyldecom1}
\end{equation}
and passing from the electric part of the Riemann tensor
$\,A_{ij}\,$ in equation (\ref{defab}) to that of the Weyl tensor
defined as
\begin{equation}
E_{ij}={}^{(5)}C_{ABCD}\,n^A n^C e^B_i e^D_j\,\,,
\label{weyldecom}
\end{equation}
we find that
\begin{equation}
A_{ij}= E_{ij}+\frac{1}{3}\,\left[{}^{(5)}R_{ij}+
\frac{1}{4}h_{ij}{}^{(5)}R -
\frac{1}{2}\,h_{ij}\left(R-K^2+K_{lm}K^{lm}\right)\right]\,\,.
\label{weylel}
\end{equation}
Using this equation one can easily relate the tensors $\,W_{ij}\,$
and $\,E_{ij}\,$. We find that
\begin{equation}
E_{ij}= W_{ij}+\kappa_5^2 \,U_{ij}\,\,, \label{ewrel}
\end{equation}
which shows that the traceless projection of the bulk
energy-momentum tensor (\ref{utrl}) determines the
difference between the electric part of the five-dimensional Weyl
tensor and the traceless tensor $\,W_{ij}\,$, constructed from the
electric part of the Riemann tensor as in equation
(\ref{wtrl}). In the particular case, where the
energy-momentum tensor of the braneworld admits the form \cite{sms1}
\begin{equation}
\tau_{ij} = - \lambda h_{ij} +S_{ij}\,, \label{branematter}
\end{equation}
where $\lambda $ is the brane tension, we can transform equation
(\ref{breq3}) into the following form
\begin{equation}
G_{ij} = -\Lambda h_{ij} + \kappa_4^2 \, S_{ij} + \kappa_5^4
\,M_{ij} - W_{ij} - 3 \kappa_5^2\,U_{ij}\,\,,
\label{breq4}
\end{equation}
where
\begin{eqnarray}
\Lambda &=& \frac{1}{2} \left(\Lambda_5 + \frac{1}{6}\kappa_5^4
\,\lambda^2 - \kappa_5^2\, P\right)\,,
\label{cosmcons}\\
\kappa_4^2 &=& \frac{1}{6}\,\kappa_5^4\,\lambda \,\,,
\label{gravconst4}\\
M_{ij} &=& - \frac{1}{4} \left[ \left(S^m_i S_{mj} -
\frac{1}{3}\,S S_{ij}\right) - \frac{1}{2}\,h_{ij}
\left(S_{lm}S^{lm} - \frac{1}{3}\,S^2\right) \right]\,\,.
\label{quademt2}
\end{eqnarray}
It is important to note that these equations do not involve any
terms determined by  a nonvanishing accerelation of normals
inherent in the ADM type foliation of spacetime. In the absence of
the energy-momentum sources in the bulk, equations (\ref{breq4})
have exactly the same form as the effective field equations
derived by Shiromizu, Maeda and Sasaki (SMS) in  \cite{sms1}.
Thus, the form of the SMS equations still continues to be held
in the ADM type non-geodesic slicing of a spacetime manifold.

We see that the effective field equations (\ref{breq4})
drastically differ from the usual four-dimensional Einstein
equations not only due to the term quadratic in the
energy-momentum tensor of the brane (\ref{quademt2}), but also due to the
curvature and energy-momentum terms from the bulk space. In other
words, the gravitational effects of the bulk  are transmitted to
the brane through a traceless second-rank tensor $\,W_{ij}\,$,
while the influence of the bulk space energy and momentum
on the brane is felt through both the normal compressive pressure
and the traceless in-brane components of the bulk energy-momentum
tensor. On these grounds, it is clear that the effective equations
on the brane are not closed, thereby there exists a continuous
exchange of energy and momentum between the brane and the bulk.

In order to demonstrate these features we appeal to equation
(\ref{constr2}) and insert into it the relation given in
(\ref{brextcur}). Then we arrive at the following equation of
"continuity"
\begin{equation}
D_m\,\tau^m_i=-2~^{(5)}J_i\,\,, \label{flux1}
\end{equation}
which manifests the existence of an energy flux onto, or away from
the brane. In the case where  $^{(5)}J_i\equiv 0$, we have
conservation equation for energy-momentum on the brane.

Next, we calculate the divergence of equation (\ref{breq3}).
Taking into account the contracted Bianchi identities
\begin{equation}
D^i G_{ij}=0 \,\,, \label{bianchi1}
\end{equation}
after some rearrangements we find that
\begin{eqnarray}
D^i \left(W_{ij}+ 3 \kappa_5^2\, U_{ij} - \frac{1}{2}\, \kappa_5^2
\,h_{ij} P \right) & = & \frac{1}{4}\,\kappa_5^4 \left[\tau^{mi}
\left(D_j \tau_{mi}-D_i\,\tau_{mj}\right) +
\frac{1}{3}\,\left(\tau^i_j - \delta^i_j \,\tau \right) D_i\,\tau
\right. \nonumber \\[2mm]
&&  \left. + 2~^{(5)}J_i \left(\tau^i_j-\frac{1}{3} \delta^i_j
\,\tau \right) \right]\, .
\label{divergence}
\end{eqnarray}
From this equation it follows that the divergence of the
corrections from the bulk space to the effective field equations
is completely determined by the distribution of
matter on the brane, though the total behaviour of these quantities must
obey the evolution equation in the bulk. In the case of a vacuum on the
brane $\,( \tau^i_j=0 )\,$, equations (\ref{breq4}),
(\ref{divergence}) and (\ref{flux1}) are reduced to the form
\begin{equation}
G_{ij} = -\Lambda h_{ij} - W_{ij} - 3 \kappa_5^2\,U_{ij}\,\,,
\label{short}
\end{equation}
\begin{equation}
D^i \left(W_{ij}+ 3 \kappa_5^2\, U_{ij} - \frac{1}{2}\, \kappa_5^2
\,h_{ij} P \right) = 0 \,\,,
\label{integr}
\end{equation}
where  equation (\ref{integr}) is of an integrability condition
on the brane. In the case of empty bulk and vanishing cosmological
constant $\,\Lambda=0\,$, one can solve equations (\ref{short})
and (\ref{integr}) using a special metric anzatz on the brane
that makes the system of equations closed \cite{roy1}.
Since the tensor $\,W_{ij}\,$ is traceless, the authors of paper
\cite{roy2} made a prescription for mapping
four-dimensional general relativity solutions with traceless
energy-momentum tensor to vacuum braneworld solutions
in five-dimensional gravity. In particular, they found the exact
solution for a static black hole on a $3$-brane with a {\it tidal}
charge arising due to gravitational influnce of the fifth dimension
on the brane. A similar approach has been used in \cite{aemir}
to construct exact solutions for rotating and charged black holes
localized on the brane. It is clear that with further assumptions
on the metric structure on a $3$-brane and on the projections
$\,P\,$ and  $\,U_{ij}\,$ of the bulk energy-momentum tensor, one can
extend the link between some solutions in general relativity and
braneworlds to include contributions from non-empty bulk space as well.

\vspace{5mm}
\section{The Evolution Equations}

As we have emphasized above, the effective gravitational field equations
(\ref{breq3}) and (\ref{breq4}) on the brane in general are not
closed and the evolution equations into the bulk are
needed to be solved for the projected bulk curvature and energy-momentum
tensors. In this section we shall derive these evolution
equations. First, we note that using equations (\ref{wtrl}) and
(\ref{atrl}) along with (\ref{excev1}), for the traceless tensor
$\,W_{ij}\,$ defined from the "electric part" of the bulk Riemann tensor
we obtain
\begin{eqnarray}
W_{ij}&=&-\frac{1}{N}\left[\left(\partial_5-
\pounds_{\vec{N}}\right) \left(K_{ij}-\frac{1}{4}\,K
h_{ij}\right)+D_i D_j N\right] -\frac{1}{2}\,K K_{ij}
+K^m_i K_{mj}\,\nonumber\\
&&+\frac{1}{4}\,h_{ij} \left(K_{lm}K^{lm} +\frac{1}{N} \Box N
\right)\,\,.
\label{wtrl1}
\end{eqnarray}
We recall that the "magnetic part",  $\,B_{jim}\,$, of the Riemann tensor
defined in equation (\ref{defab}) is not traceless. It may be useful
to construct from this quantity its traceless counterpart.
This is given by
\begin{equation}
M_{jim} = B_{jim} + \frac{2}{3}\,\kappa_5^2 \,h_{j[i} J_{m]} \,,
\label{mtrl}
\end{equation}
where we have used equation (\ref{constr2}). It turns out that
this quantity  is exactly the same as the "magnetic part" of the
Weyl tensor introduced in \cite{sms1}. We now need to establish the
equations governing the evolution of the quantities  $\,W_{ij}\,$
and $\,M_{jim}\,$ into the bulk. For this purpose, it is convenient
to start with equation (\ref{breq2}), which can also be rewritten
in the form
\begin{equation}
R_{ij}= \frac{1}{2} h_{ij}\, \left(\Lambda_5 - \kappa_5^2\,P
\right) - W_{ij} -3 \kappa_5^2 \,U_{ij} - K_i^m K_{mj}+K
K_{ij}\,\,. \label{breq5}
\end{equation}
Acting on both sides of this equation by the operator
$\,\partial_5- \pounds_{\vec{N}} \,$ one can express the evolution
of $\,W_{ij}\,$ in terms of the evolution of the braneworld Ricci tensor
$\,R_{ij}\,$ which, in turn, can be determined through the five-dimensional
Bianchi identities
\begin{equation}
\nabla_{[A}\ ^{(5)} R_{BC]DE}=0\,.
\label{bianchi5d}
\end{equation}
Using equations (\ref{redchr}) and (\ref{redriemann})
along with (\ref{empart}) one can decompose (\ref{bianchi5d})
to obtain the following set of identities
\begin{equation}
D_{[i} R_{jl]ms} = 0 \,, \label{bianchi4d}
\end{equation}
\begin{equation}
D_{[m} B_{|j|lk]} + K^i_{[m} R_{kl]ij} = 0\,, \label{cyclicid}
\end{equation}
\begin{eqnarray}
\frac{1}{N}\left(\partial_5 - \pounds_{\vec{N}} \right) B_{jml} &
- & 2 \left(D_{[m} A_{l]j} + K^i _{[m}
B_{l]ji} - A_{j[l} a _{m]} \right)\,\nonumber\\
& - & K^i _j \,B_{iml} - \  ^{(5)}R_{ijml}\,a^i=0\,, \label{eB}
\end{eqnarray}
\begin{eqnarray}
\frac{1}{2N}\left(\partial_5 - \pounds_{\vec{N}} \right) R_{ijlm}
& + & D _{[l} B_{m]ji} + a_{[i} B_{j]lm} + a_{[l} B_{m]ij}
- 2 K _{[i|[l} D _{m]|} a_{j]}\,\nonumber\\
& + & R _{lms[i} K ^s _{j]} - 2 a _{[i} K_{j][l} a _{m]} = 0\,,
\label{eriemann}
\end{eqnarray}
here and in what follows the vertical bars at both sides of an
index, or indices indicate that the indices are omitted from
(anti)-symmetrization.

Taking the trace of equation (\ref{eriemann}) and comparing the
result with equation (\ref{breq5}) after acting on it by the
operator $\,\partial_5- \pounds_{\vec{N}} \,$, we obtain the
following evolution equation for the projected bulk curvature
and energy-momentum tensors involved in the effecive field
equations on the brane
\begin{eqnarray}
\frac{1}{N} \left( \partial_5 - \pounds_{\vec{N}} \right)&&\! \!
\! \! \left(W _{ij} + 3 \kappa_5^2\,U _{ij} + \frac{1}{2} \kappa
_5 ^2 \,P h _{ij} \right)
 = \left( D ^m - 2 a ^m \right) M _{(ij)m} + 3 \left( K^m _{(i} \,W
_{j)m} + \kappa_5^2\,K ^m _{(i} \,U _{j)m} \right)\,\nonumber\\ &&-\left[
K W _{ij} + \left( K _{ij} K _{lm} - K _{im} K _{jl} \right) K
^{lm} - R _{limj} K ^{lm} - \frac{1}{6}\, \Lambda_5 \left( K _{ij} -
h _{ij} K
\right) \right]\,\nonumber\\
&& + \frac{2}{3}\, \kappa _5 ^2 \left[ 2\left( J _{(i} a _{j)} +
\frac{1}{2} h _{ij} J _m a ^m\right) - \left( D _{(i} J _{j)} +
\frac{1}{2} h _{ij}\, D_m J ^m\right) \right]
\nonumber\\
&&+ \frac{1}{4}\, \kappa _5 ^2 \left[\frac{2}{3}\,\ ^{(5)}T
\left(K _{ij} + \frac{1}{2} h _{ij} K \right) - P h _{ij} K
\right] \,\, \label{evolbulk}
\end{eqnarray}
In obtaining  this equation we have also used Bianchi identities
given by equations (\ref{bianchi4d}) and (\ref{cyclicid}). Taking
into account the expression (\ref{mtrl}) in equation (\ref{eB}) we
we arrive at the evolution equation for the traceless "magnetic" part of
the Weyl tensor
\newpage
\begin{eqnarray}
\frac{1}{N} \left(\partial_5 - \pounds_{\vec{N}} \right) M _{jml}
&=& 2 \left( D _{[m} W_{l]j} + K ^i _{[m} M _{l]ji} + W _{j[m}
a_{l]}\right) + K ^i _j M _{iml} + a ^i \ ^{(5)} R _{ijml} \,
\nonumber\\ & - & \frac{2}{3}\, \kappa _5 ^2 \left[ \frac{1}{N}
h_{j[l} \left(
\partial_5 - \pounds_{\vec{N}} \right) J_{m]} + 2 \left( K ^i
_{[l}\, h _{m][i} J _{j]} + K _{j[l} J _{m]} \right) + K ^i _j
\,h_{i[m} J _{l]}\right]\,\nonumber\\ &+& \frac{1}{2}\, \kappa _5 ^2
\,h_{j[l} D _{m]} \left(P - \frac{1}{3} \ ^{(5)} T \right) -
\frac{1}{2} \left[ \kappa _5 ^2 \left(P - \frac{1}{3} \ ^{(5)}
T \right)+ \frac{2}{3} \Lambda_5 \right] h _{j[l} a _{m]} \,
\label{evolM}
\end{eqnarray}
The equations (\ref{wtrl1}), (\ref{evolbulk}) and (\ref{evolM})
generalize the corresponding equations obtained
in \cite{sms1} to the case of non-geodesic ADM type slicing of the
bulk spacetime with nonvanishing matter fields.
We note that here we deal with the tensor $\,W_{ij}\,$ which looks
more simpler and it coincides with $\,E_{ij}\,$ when the bulk is empty.
We see that the evolution equations are significantly modified due to
the nonvanishing acceleration of spacelike normals to the brane surface,
as well as the presence of the other fields in the bulk. Clearly,
these equations are needed to be solved with appropriate
boundary  conditions on the brane. They are given by equation
(\ref{divergence}) and also by
\begin{equation}
\left[M _{jml}\right] = -2\, \kappa _5 ^2 D _{[l} \left( \tau
_{m]j} - \frac{1}{3}\, h _{m]j}\, \tau \right) - \frac{2}{3}\,
\kappa _5 ^2\, D_i \left(h _{l[j} \tau ^i _{m]}\right)\,,
\label{jumpM}
\end{equation}
which follows from the Israel junction conditions (\ref{jcisrael})
along with taking into account equation (\ref{constr2}). In the
case of a vacuum in the bulk $\,^{(5)}T_{AB}\equiv 0 \,$ , the above
equations are simplified to have the form
\begin{eqnarray}
\frac{1}{N}\left( \partial_5 - \pounds_{\vec{N}} \right) W _{ij}
&=& \left(D ^m - 2 a ^m \right) M _{(ij)m} + 3 K ^m _{(i} W_{j)m}
- K W _{ij} + R _{limj} K ^{lm}\,\nonumber\\ &&+ \left(K _{im} K
_{jl} - K _{ij} K _{lm}\right)K ^{lm} + \frac{1}{6}\, \Lambda_5 \left(
K _{ij} - h _{ij} K \right)\,, \label{empty1}
\end{eqnarray}
\begin{eqnarray}
\frac{1}{N}\left( \partial_5 - \pounds_{\vec{N}} \right) M _{jml}
&=& 2 \left( D _{[m} W _{l]j} +K ^i _{[m} M _{l]ji} + W _{j[m} a
_{l]} \right) + K ^i _j M _{iml} \,\nonumber\\ &&+ a ^i \ ^{(5)} R
_{ijml} - \frac{1}{3}\, \Lambda_5\, h _{j[l} a _{m]}\,. \label{empty}
\end{eqnarray}
When passing to the Gausssian normal coordinates these equations
are in agreement with those obtained in \cite{sms1}. Thus, the
equations governing the evolution of quantities $\,W_{ij}\,$  and
$\,M_{jml}\,$ into the fifth dimension are the desired supplements
to the effective field equations on the $3$-brane given in
(\ref{constr1}), (\ref{constr2}) and (\ref{breq4}) making the full
system of equations closed.

\section{Conclusion}
We have generalized the effective gravitational field equations on
and off a $3$-brane with $\,Z_2\,$ symmetry to the case where
there exists an arbitrary energy-momentum tensor in the bulk. We
have adopted a general coordinate setting introducing the lapse
scalar field and the shift vector field, in accordance with their
corresponding counterparts in ADM type slicing of a spacetime in
four-dimensional general relativity. The use of the ADM type
coordinates provides complete freedom in the slicing of a five
dimensional bulk spacetime by pushing the timelike hypersurfaces
forward in the fifth dimension at arbitrary rates in different
positions. We have shown that the form of the on-brane effective
gravitational field equations obtained earlier by Shiromizu, Maeda
and Sasaki in the framework of the Gaussian normal coordinates
continues to be held in the ADM type coordinate setting,
however, the evolution equations into the bulk are significantly
changed due to both the acceleration of normals inherent in the ADM
approach and the presence of other bulk fields. We have also shown
that the gravitational effects of the fifth dimension on a $Z_2$-symmetric
$3$-brane may be described by means of a traceless second-rank
tensor, properly constructed from the "electric" part of the
Riemann tensor from the bulk space.

It should be emphasized that the formalism used in this paper might be very
useful in studying gravitational perturbations in braneworld gravity
and cosmology. The use of a general lapse function and shift vector
may provide one with a new gauge, thereby taking into account the "brane
-bending" type effects consistently. In particular, it would be interesting
to explore the effects of the universal aspects of linearized gravity
in a $3$-braneworld, solving the on-brane and the off-brane
equations for corresponding metric fluctuations.

\section{Acknowledgments}

We would like to thank R. Maartens for discussions and valuable comments.
One of us (A.N.) also thanks M. Sasaki for useful discussions
at an early stage of this work.



\begin{thebibliography}{99}

\bibitem{pol} J. Polchinski, TASI Lectures on D-branes, hep-th/9611050
\bibitem{hw}  P. Horava and E. Witten, Nucl. Phys. B {\bf 460}, 506 (1996).
\bibitem{w} E. Witten, Nucl. Phys. B {\bf 471}, 135 (1996).
\bibitem{lukas} A. Lukas, B.A. Ovrut, K.S. Stelle, and D. Waldram,
Phys. Rev. D {\bf 59}, 086001 (1999).
\bibitem{rsh} V. A. Rubakov and M. E. Shaposhnikov, Phys. Lett. B
 {\bf 125}, 136 (1983).
\bibitem{aka} K. Akama, {\it Lecture Notes in Physics},
 {\bf 176}, 267 (1982) hep-ph/0001113.
\bibitem{vis1} M. Visser, Phys. Lett. B {\bf 159}, 22 (1985).
\bibitem{gw}  G. W. Gibbons and D. L. Wiltshire, Nucl. Phys. B {\bf 287},
717 (1987).
\bibitem{ADD} N. Arkani-Hamed, S. Dimopoulos, and G. Dvali, Phys. Lett. B
{\bf 429}, 263 (1998); I. Antoniadis, N. Arkani-Hamed, S.
Dimopoulos, and G. Dvali, Phys. Lett. B {\bf 436}, 257 (1998).
\bibitem{emp1} R. Emparan, M. Masip, and R. Rattazzi,
Phys. Rev. D {\bf 65}, 064023 (2002)
\bibitem{gidt} S. B. Giddings and S. Thomas, Phys. Rev. D {\bf 65},
056010 (2002).
\bibitem{dl} S. Dimopoulos and G. Landsberg, Phys. Rev. Lett. {\bf 87},
161602 (2001).
\bibitem{rs1} L. Randall and R. Sundrum, Phys. Rev. Lett. {\bf 83},
3370 (1999).
\bibitem{rs2} L. Randall and R. Sundrum, Phys. Rev. Lett. {\bf 83},
4960 (1999).
\bibitem{csaki} C. Csaki, {\it TASI Lectures on Extra
Dimensionsand Branes}  hep-ph/0308112.
\bibitem{roy} R. Maartens, {\it  Brane-World Gravity}  gr-qc/0312059.
\bibitem{gabad} G. Gabadadze, {\it ICTP Lectures on Large Extra
Dimensions}  hep-ph/0308112.
\bibitem{rubakov} V. Rubakov, Phys. Usp. {\bf 44}, 871 (2001);
Usp. Fiz. Nauk {\bf 171}, 913 (2001).
\bibitem{gt} J. Garrigia and T. Tanaka, Phys. Rev. Lett. {\bf 84},
2778 (2000).
\bibitem{gkr} S. B. Giddings, E. Katz and L. Randall, J. High Energy Phys.
{\bf 0003}, 23 (2000).
\bibitem{chgw} A. Chamblin and G.W. Gibbons, Phys. Rev. Lett. {\bf 84},
1090, (2000)
\bibitem{sms1} T. Shiromizu, K. Maeda, and M. Sasaki, Phys. Rev. D {\bf 62},
024012 (2000).
\bibitem{sms2} M. Sasaki, T. Shiromizu and K. Maeda,
Phys. Rev. D {\bf 62}, 024008 (2000).
\bibitem{mw} K. Maeda and D. Wands, Phys. Rev. D {\bf 62},
124009 (2000).
\bibitem{mt} K. Maeda and  T. Torii, Phys. Rev. D {\bf 69},
024002 (2004).
\bibitem{mtw}  C. W. Misner, K. S. Thorne, and J. A. Wheeler,
{\em Gravitation}, W. H. Freeman and Company (1973).
\bibitem{adm} R. Arnowitt, S. Deser, and C.W. Misner, In {\it Gravitation:
an introduction to current research}, L. Witten, ed. Wiley, New
York, p.227 (1962); gr-qc/0405109.
\bibitem{gh} G.W. Gibbons and S.W. Hawking, Phys. Rev. D {\bf 12},
2752 (1977).
\bibitem{Israel} W. Israel, Nuovo Cimento  {\bf 44B}, 1 (1966); Errata-ibid
 {\bf 48B}, 463 (1967).
\bibitem{eric} E. Poisson, {\it A Relativist's Toolkit: The Mathematics of
Black Hole Mechanics}, Cambridge University Press  (2004)
\bibitem{landau} L.D. Landau and E.M. Lifshitz, {\it The Classical Theory
of Fields}, Pergamon Press (1975).
\bibitem{roy1} R. Maartens, Phys. Rev. D {\bf 62}, 084023 (2000).
\bibitem{roy2} N. Dadhich, R. Maartens, P. Papadopoulos, and V. Rezania,
 Phys. Lett. B {\bf 487}, 1 (2000).
\bibitem{aemir} A.N. Aliev and A.E. Gumrukcuoglu, in preparation
(2004)
\end{thebibliography}
\end{document}